\documentclass{article}
\usepackage{amsmath}

\usepackage[numbers, super]{natbib}
\usepackage{threeparttable}

\makeatletter
\renewcommand\@biblabel[1]{#1.}
\makeatother
\usepackage{float}
\usepackage{subcaption}
\usepackage{graphicx}
\usepackage{enumitem} 
\usepackage[utf8]{inputenc} 
\usepackage[T1]{fontenc}    
\usepackage{hyperref}       
\usepackage{url}            
\usepackage{booktabs}       
\usepackage{amsfonts}       
\usepackage{nicefrac}       
\usepackage{microtype}      
\usepackage{xcolor}         

\usepackage[final]{neurips_2023}

\title{GitHub Stargazers | Building Graph- and Edge-level Prediction Algorithms for Developer Social Networks}

\author{%
  Karishma Thakrar \\
  Georgia Institute of Technology \\
  \AND
  Aniket Chauhan \\
  Georgia Institute of Technology \\
}

\begin{document}

\maketitle


\renewcommand{\thefootnote}{\fnsymbol{footnote}}
\footnotetext[1]{This architecture was first created in November 2023 and the paper was later published on arXiv in January 2025. The code is available at \url{https://github.com/karishmathakrar/stargazers}.}
\renewcommand{\thefootnote}{\arabic{footnote}}

\begin{abstract}

Analyzing social networks formed by developers provides valuable insights for market segmentation, trend analysis, and community engagement. In this study, we explore the GitHub Stargazers dataset to classify developer communities and predict potential collaborations using graph neural networks (GNNs). By modeling 12,725 developer networks, we segment communities based on their focus on web development or machine learning repositories, leveraging graph attributes and node embeddings. Furthermore, we propose an edge-level recommendation algorithm that predicts new connections between developers using similarity measures. Our experimental results demonstrate the effectiveness of our approach in accurately segmenting communities and improving connection predictions, offering valuable insights for understanding open-source developer networks.

\end{abstract}

\section{Introduction}

Effective market segmentation provides substantial advantages, as evidenced by a study conducted by Bain \& Company.\cite{carpenter2018market} According to this study, 81\% of executives consider segmentation crucial for profit growth, with organizations employing effective segmentation strategies reporting a 10\% higher profit over five years compared to those with less effective approaches. In the technology sector, particularly in the analysis of social networks formed by developers, segmentation plays a critical role. This study leverages the GitHub Stargazers dataset\cite{karateclub} from the Stanford Network Analysis Project (SNAP) to explore this phenomenon.

The dataset, consisting of 12,725 graphs with 10 to 957 users each, presents challenges due to its unstructured nature and the complexity of interactions among various developer communities. To address these challenges, this work applies predictive and scalable graph neural network (GNN) models to segment the social networks into communities focused on either web development or machine learning. This segmentation serves as a foundation for market and trend analysis within the tech sector, providing insights into technological advancements and emerging areas of developer interest.

Understanding network evolution is essential for market research, as it reveals trends in community development and shifts in user interests, enabling businesses to formulate targeted strategies that align with evolving community needs. This study examines these dynamics by analyzing and modeling node features and embeddings. The methodology further aims to identify potential new connections between nodes and develop an advanced recommendation algorithm, enhancing user experience and fostering engagement within the network.

The primary objectives of this research are:

\begin{itemize}
\item To model and classify developer networks within the dataset, providing a structured segmentation of communities based on their characteristics as either web development or machine learning.
\item To develop a recommendation algorithm that predicts potential node connections within the network.
\end{itemize}

By achieving these objectives, this study contributes to a deeper understanding of open-source communities, capturing the nuanced dynamics and interactions among developers.

\section{Data Sources \& Preprocessing}

\subsection{Data Source}
The \textbf{GitHub Stargazers} dataset from the Stanford Network Analysis Project (SNAP), curated by Jure Leskovec, was utilized for this study.\cite{karateclub}
\begin{itemize}
    \item \textbf{Dataset link}: \url{https://snap.stanford.edu/data/github_stargazers.html}
\end{itemize}

The dataset consists of 12,725 individual graphs, each representing a network of developers who have starred repositories on GitHub. Each network contains between 10 and 957 nodes, where nodes represent users, and edges represent undirected follower relationships. The dataset does not include additional descriptive features for nodes or edges. However, each network is labeled according to its primary association with either web development or machine learning repositories, facilitating the classification of developer communities.

\subsection{Data Preprocessing for Exploratory Data Analysis (EDA)}
To analyze the structural properties of these networks, the dataset was first converted into graph objects using \texttt{NetworkX},\cite{hagberg2008networkx} a Python library for handling graph structures. This representation enabled a systematic examination of relationships and interactions within the networks. Several key network statistics were computed to characterize the dataset:

\begin{itemize}
    \item \textbf{Node Count}: The total number of developers (nodes) in each network.
    \item \textbf{Edge Count}: The total number of connections (edges) between developers.
    \item \textbf{Average Degree}: The mean number of connections per node, calculated by summing all node connections and dividing by the total number of nodes.
    \item \textbf{Density}: A measure of how connected the network is, computed as the ratio of actual edges to the maximum possible number of edges.
\end{itemize}

These metrics provided insights into network connectivity patterns and served as a foundation for further analysis.\cite{grando2019mlcentrality}

\subsection{Data Preprocessing for Modeling}
For graph-based modeling, the dataset was processed using \texttt{Deep Graph Library (DGL)}, which supports efficient implementation of Graph Neural Networks.\cite{dgl2023} Each network was converted into a DGL graph object, incorporating both basic network properties and additional structural features to enhance predictive modeling. The extracted features included:

\begin{itemize}
    \item \textbf{Degree}: The number of direct connections each node has.
    \item \textbf{Clustering Coefficient}: The likelihood of a node’s neighbors being connected, indicating local network cohesiveness.
    \item \textbf{Betweenness Centrality}: A measure of how frequently a node appears on the shortest paths between other nodes, capturing its influence in network communication.
    \item \textbf{Closeness Centrality}: An indicator of how efficiently a node can reach all other nodes, based on the average shortest path distance.
    \item \textbf{PageRank}: A metric assessing node importance based on the quantity and quality of its connections, originally developed for ranking web pages.
\end{itemize}

Following feature extraction, labels were converted into \texttt{PyTorch} tensors, and the dataset was partitioned into training (60\%), validation (20\%), and testing (20\%) sets. Dataloaders were employed to process batches of 32 samples, optimizing computational efficiency during model training.

This structured preprocessing ensured that the dataset was effectively prepared for both exploratory analysis and predictive modeling, enabling robust classification and recommendation tasks.

\section{Exploratory Data Analysis}

\subsection{Node and Edge Distribution}
\begin{figure}[H]
    \centering
    \includegraphics[width=1\linewidth]{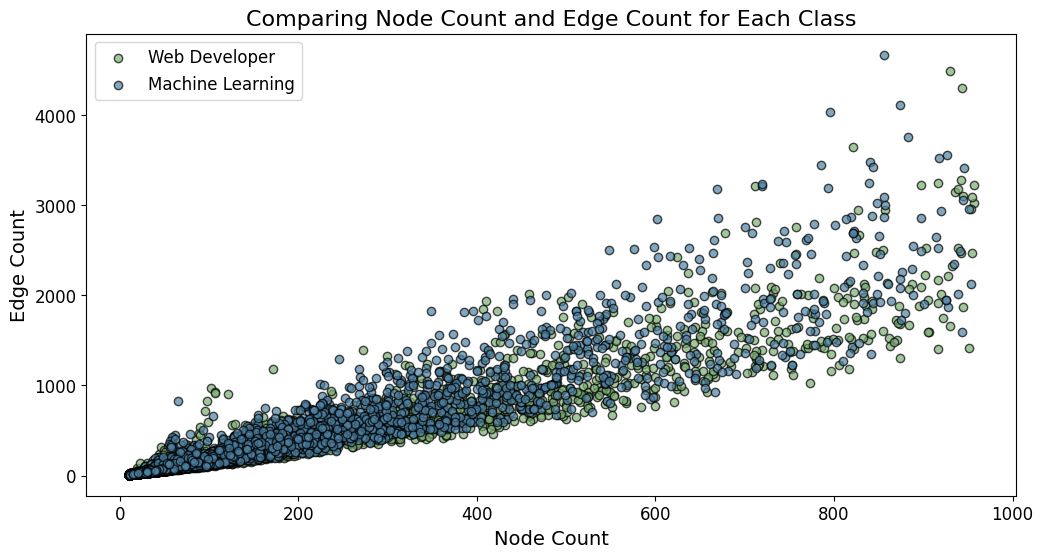}
    \caption{Comparison of Node Count and Edge Count for Each Class.}
    \label{fig:scatter_nodes_edges}
\end{figure}

To understand the distribution of the dataset, a scatter plot comparing node count and edge count across the two categories was examined (see Figure~\ref{fig:scatter_nodes_edges}). As expected, an increase in node count was generally accompanied by a corresponding increase in edge count. While no distinct separation between web development and machine learning networks was observed, several machine learning networks exhibited a notably higher edge-to-node ratio compared to web development networks. Specifically, among the densest networks in each category, machine learning networks had an average of 5.5 edges per node, whereas web development networks had a slightly lower average of 4.8 edges per node.  

\newpage

\subsection{Correlation Analysis of Network Features}
\begin{figure}[H]
    \centering
    \includegraphics[width=0.85\linewidth]{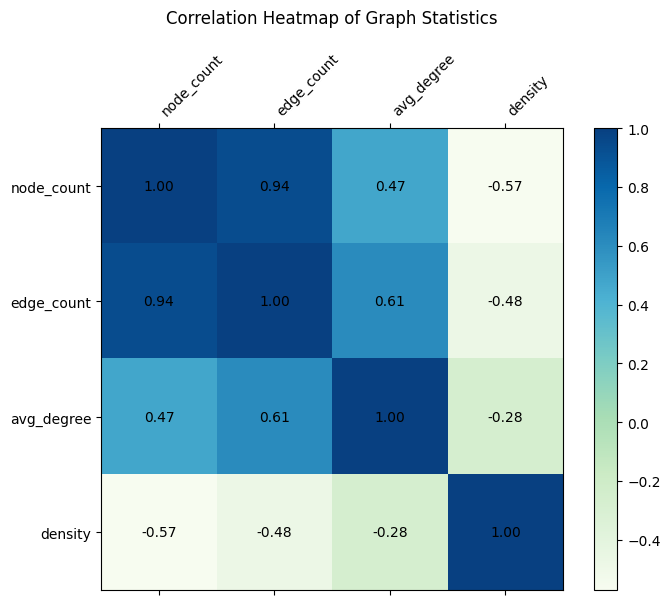}
    \caption{Correlation Heatmap of Graph Statistics.}
    \label{fig:sample}
\end{figure}

Further analysis was conducted to investigate relationships between different network features. A correlation heatmap (see Figure~\ref{fig:sample}) revealed a strong correlation of 0.94 between node count and edge count, reinforcing the notion that larger graphs tend to have greater absolute connections. Additionally, node count and edge count exhibited a moderate negative correlation with density (-0.57 and -0.48, respectively), indicating that as networks increase in size, their relative connectivity decreases. A positive correlation of 0.61 was also observed between edge count and average degree, which aligns with the expectation that networks with more edges generally have higher average degrees.  

\newpage

\subsection{Comparison of Web Development and Machine Learning Networks}
When comparing key structural metrics between web development and machine learning networks, machine learning networks were observed to have, on average, more nodes, more edges, higher average degree, and higher density than web development networks, with very few exceptions.





\begin{figure}[H]
    \centering
    \begin{subfigure}{0.49\linewidth}
        \centering
        \includegraphics[width=\linewidth]{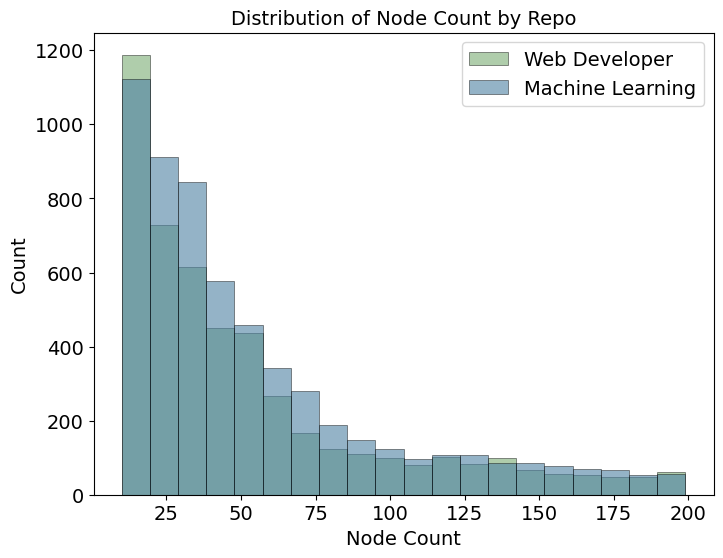}
        \caption{Node Count}
        \label{fig:network1}
    \end{subfigure}
    \hfill
    \begin{subfigure}{0.49\linewidth}
        \centering
        \includegraphics[width=\linewidth]{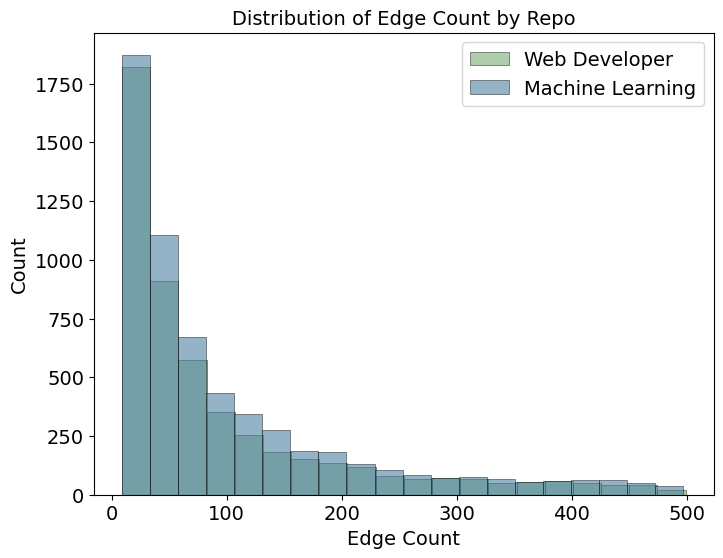}
        \caption{Edge Count}
        \label{fig:network2}
    \end{subfigure}
    \vskip\baselineskip
    \begin{subfigure}{0.49\linewidth}
        \centering
        \includegraphics[width=\linewidth]{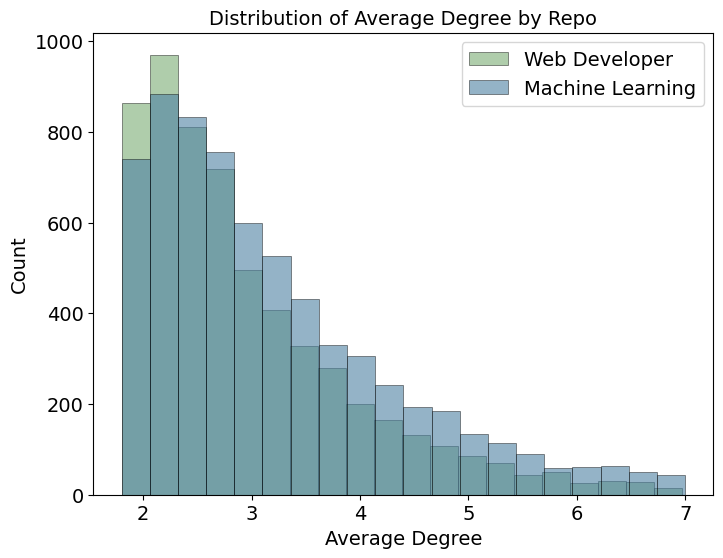}
        \caption{Average Degree}
        \label{fig:network3}
    \end{subfigure}
    \hfill
    \begin{subfigure}{0.49\linewidth}
        \centering
        \includegraphics[width=\linewidth]{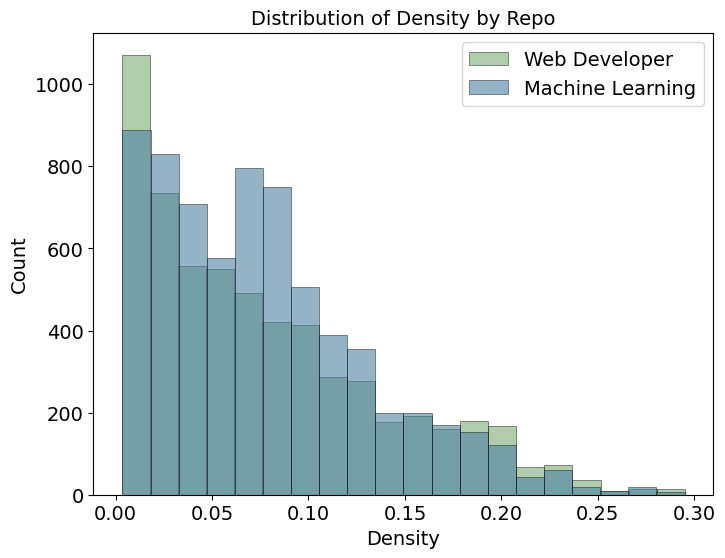}
        \caption{Density}
        \label{fig:network4}
    \end{subfigure}
    \caption{Comparing Underlying Network Structures for Each Class.}
    \label{fig:combined_network_structures}
\end{figure}

\newpage

\subsection{Visualization of Individual Networks}
To further examine the structural differences between the two network types, individual networks were visualized (see Figure 4). Web development networks exhibited sparser connectivity, with a greater presence of isolated nodes, whereas machine learning networks were denser, containing more connections between nodes. These differences align with expectations, as machine learning repositories may attract a more interconnected user base due to shared research interests and collaborations.



\begin{figure}[H]
    \centering
    \begin{subfigure}{0.7\linewidth}
        \centering
        \includegraphics[width=\linewidth]{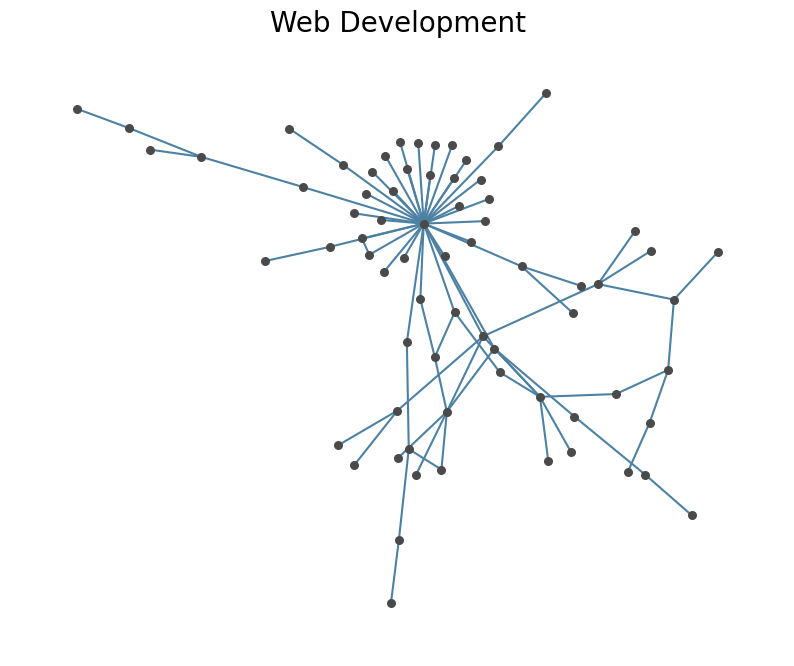}
        \label{fig:web_dev}
    \end{subfigure}
    \hfill
    \begin{subfigure}{0.7\linewidth}
        \centering
        \includegraphics[width=\linewidth]{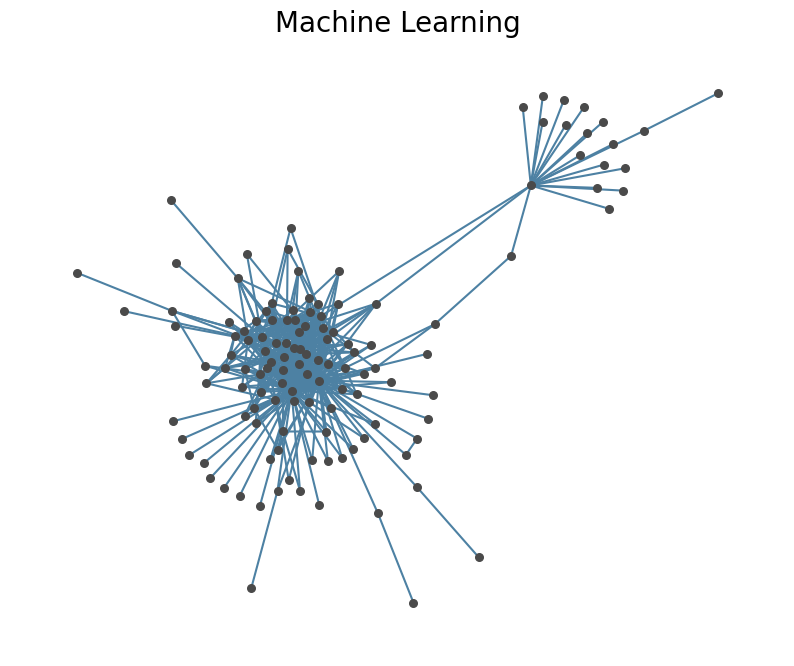}
        \label{fig:ml_network}
    \end{subfigure}
    \caption{Example Networks in Web Development and Machine Learning Demonstrating Differences in Connectivity.}
    \label{fig:combined_networks}
\end{figure}

\section{Methodology}

This study involved the development and evaluation of a recommendation algorithm using node feature learning and dot product calculations between node embeddings to predict potential new connections within developer networks. A variety of metrics were utilized to iteratively assess model performance, including training and validation curves, Receiver Operating Characteristic (ROC) curves, and Area Under the Curve (AUC).

\subsection{Graph Classification}

\subsubsection{Modeling}
Graph Convolutional Networks (GCNs) were implemented using the Deep Graph Library to develop a classification algorithm. These networks leverage message passing, allowing each node to aggregate features from its neighbors and refine its representation. The transformation applied by each graph convolutional layer is given by:

\begin{equation}
H^{(l+1)} = \sigma(\tilde{D}^{-\frac{1}{2}} \tilde{A} \tilde{D}^{-\frac{1}{2}} H^{(l)} W^{(l)})
\end{equation}

where $H^{(l)}$ and $H^{(l+1)}$ are the feature matrices of the $l^{th}$ and $(l+1)^{th}$ layers, respectively, $W^{(l)}$ is the learnable weight matrix, $\sigma$ is a non-linear activation function such as ReLU, and $\tilde{A}$ and $\tilde{D}$ represent the adjacency and degree matrices, respectively. This formulation enables information propagation across the network, allowing the model to capture both local and global graph structures.

The classification model was designed to incorporate five fundamental node features, which were integrated with graph topology through the layered architecture of the GCN. The Adam optimizer was employed for parameter updates, while Binary Cross Entropy loss guided the training process. Several architectural variations were explored, refining the network's depth and processing layers to improve classification performance.

\subsubsection{Classifier Architectures}
To assess the impact of different architectural choices, four classifier configurations were evaluated. Each variant explored different combinations of GCN layers and auxiliary processing techniques to optimize feature extraction and representation learning.

\begin{itemize}
    \item \textbf{Classifier 1}: Composed of two GCN layers and two fully connected layers, utilizing ReLU activation in the hidden layers and a sigmoid activation in the final layer. This served as a baseline to assess the influence of GCNs on classification performance.
    \item \textbf{Classifier 2}: Introduced additional complexity with three GCN layers, a sort pooling layer, two 1D convolutional layers, a fully connected layer, and a dropout layer prior to the final output. The sort pooling layer transformed variable-sized graph representations into a fixed-size format, facilitating improved learning and feature extraction.\cite{zhang2018graphclassification}
    \item \textbf{Classifier 3}: Similar in structure to Classifier 2 but with two GCN layers instead of three. This variation aimed to determine the impact of reducing GCN depth on feature learning effectiveness.
    \item \textbf{Classifier 4}: A simplified model with three GCN layers and two fully connected layers, omitting sort pooling and convolutional components. Despite its reduced complexity, this architecture ultimately demonstrated the highest performance.
\end{itemize}

The incremental modifications in Classifiers 2 and 3, such as the addition of sort pooling and dropout layers, were intended to enhance representation learning. However, the superior performance of Classifier 4 suggested that a less complex model yielded improved generalization and reduced overfitting.





\begin{figure}[H]
    \centering
    \begin{subfigure}{0.55\linewidth}
        \centering
        \includegraphics[width=\linewidth]{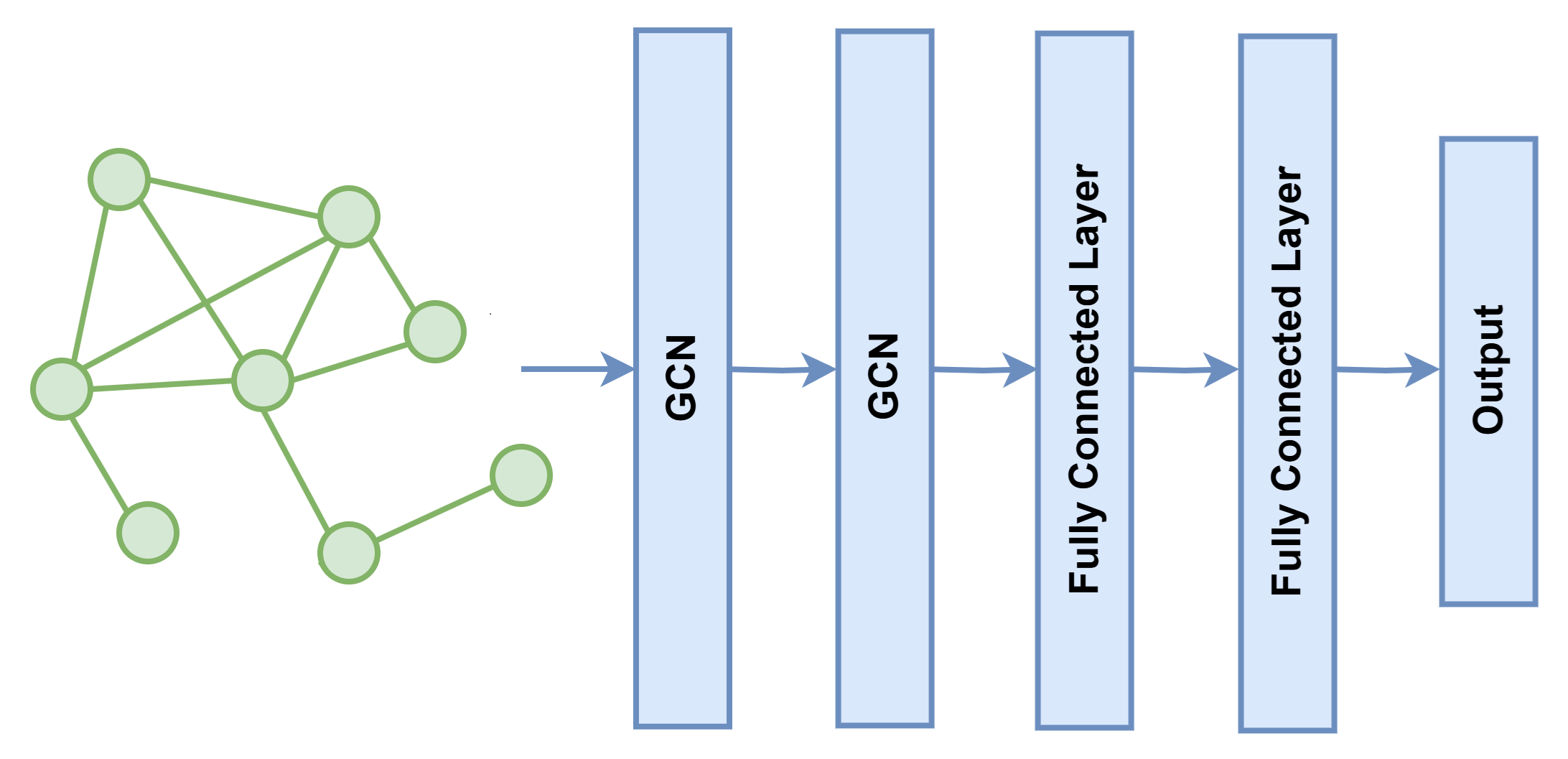}
        \caption{Classifier 1}
        \label{fig:gcn1}
    \end{subfigure}
    \hfill
    \begin{subfigure}{0.85\linewidth}
        \centering
        \includegraphics[width=\linewidth]{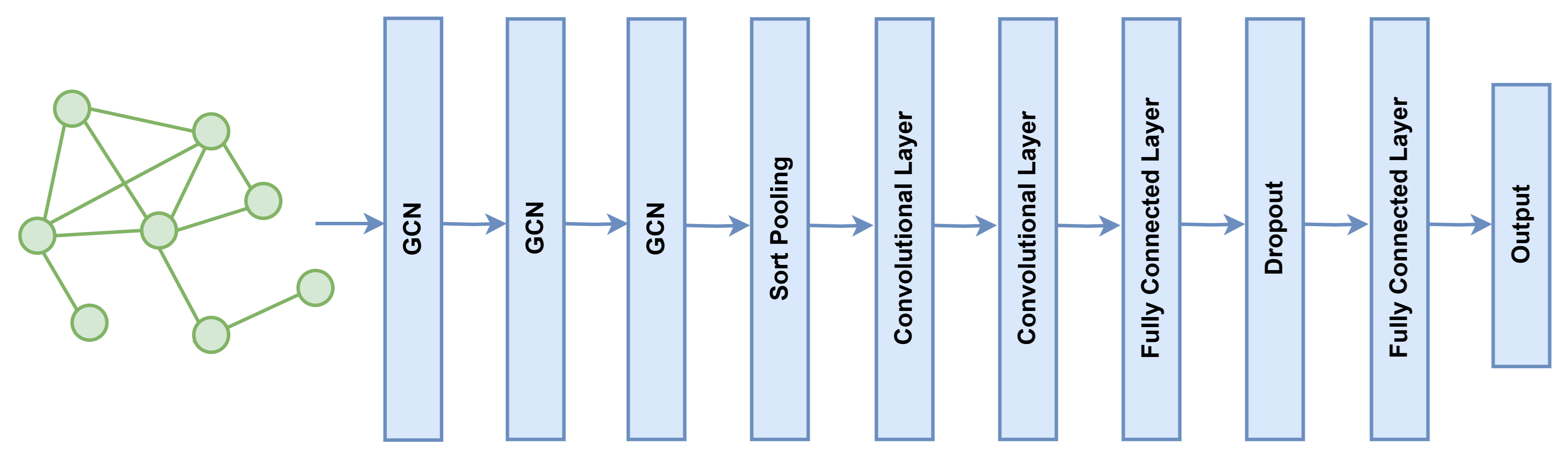}
        \caption{Classifier 2}
        \label{fig:gcn2}
    \end{subfigure}
    \vfill
    \begin{subfigure}{0.78\linewidth}
        \centering
        \includegraphics[width=\linewidth]{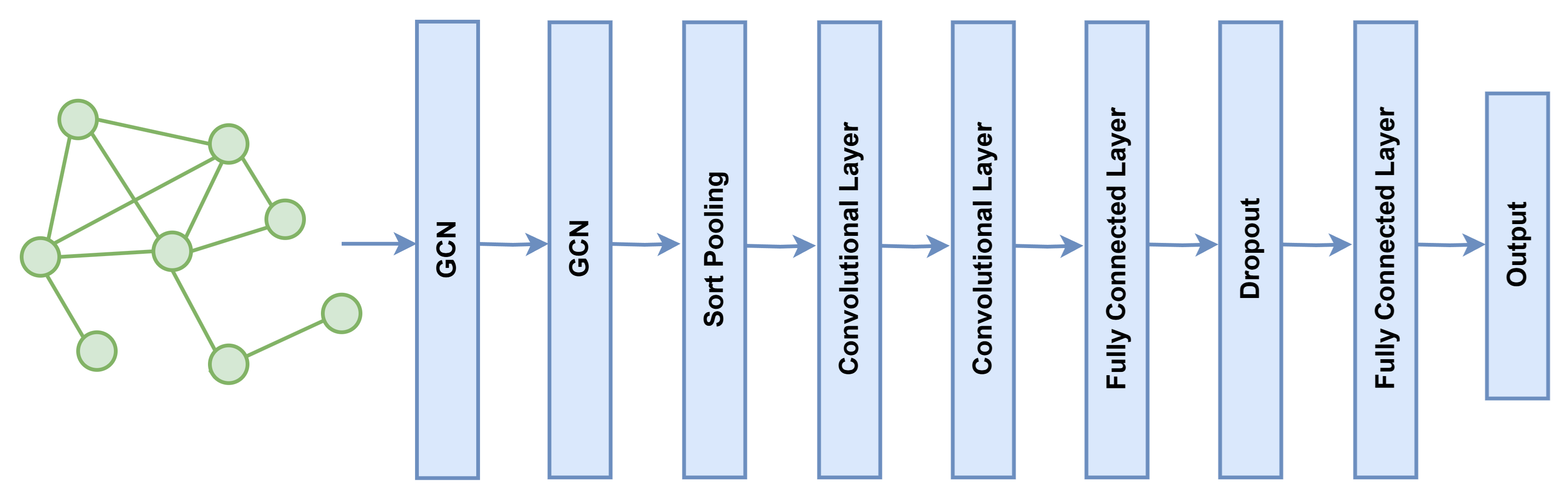}
        \caption{Classifier 3}
        \label{fig:gcn3}
    \end{subfigure}
    \hfill
    \begin{subfigure}{0.55\linewidth}
        \centering
        \includegraphics[width=\linewidth]{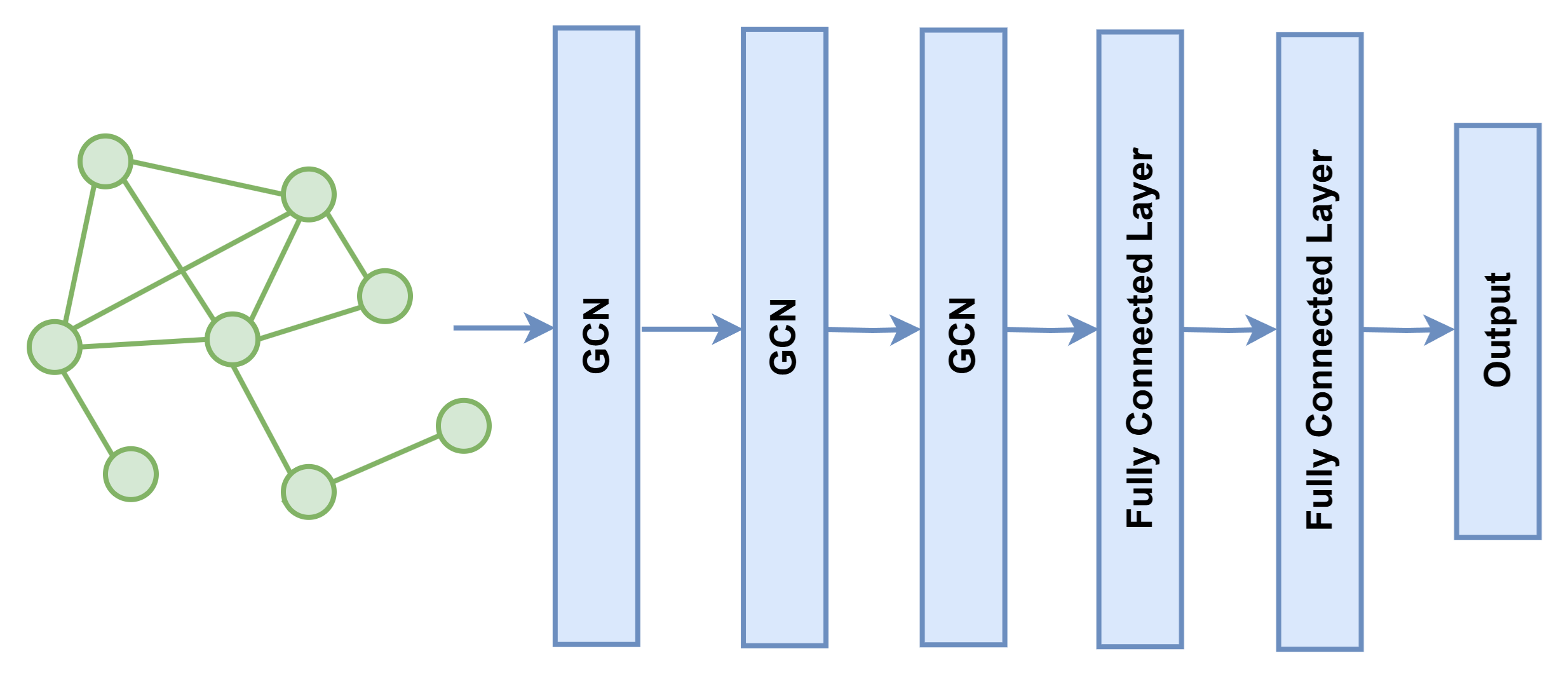}
        \caption{Classifier 4}
        \label{fig:gcn4}
    \end{subfigure}
    \caption{Comparison of Graph Convolutional Network Classifiers.}
    \label{fig:classifier_comparison}
\end{figure}

To further refine classification performance, node embeddings from the final GCN layer of Classifier 4 were extracted and utilized as input for a Random Forest classifier. This hybrid approach proved effective, as it leveraged the high-level representations generated by the GCN while benefiting from the ensemble nature of Random Forests. This combination mitigated overfitting and enhanced generalization, ultimately resulting in a more robust classification model.

\newpage



\subsection{Edge-level Recommendation System}

A graph-based recommendation system was developed to perform link prediction, identifying potential new connections within the networks. The approach leveraged Graph Sample and Aggregation (GraphSAGE) to learn node features and generate meaningful embeddings. Through its aggregation mechanism, GraphSAGE transformed the initial 5-dimensional feature space into a more expressive 100-dimensional representation per node, allowing for more effective learning of structural relationships.

To estimate the likelihood of connections, a dot product was applied between node embeddings, generating scores for both existing (positive) and non-existent (negative) edges. A margin loss function was employed to maximize the scores of positive edges while minimizing those of negative edges, enhancing the model’s ability to distinguish between true and potential connections. Additionally, cosine similarity was incorporated to refine recommendations, identifying highly similar nodes likely to form new connections.\cite{darshan2022cosine}

This edge-level recommendation framework provided valuable insights into the evolving dynamics of developer networks, capturing patterns of interaction and potential collaborations within open-source communities.

\section{Results}

\subsection{Graph Classification}

\begin{figure}[H]
    \centering
    \begin{subfigure}{0.49\linewidth}
        \centering
        \includegraphics[width=\linewidth]{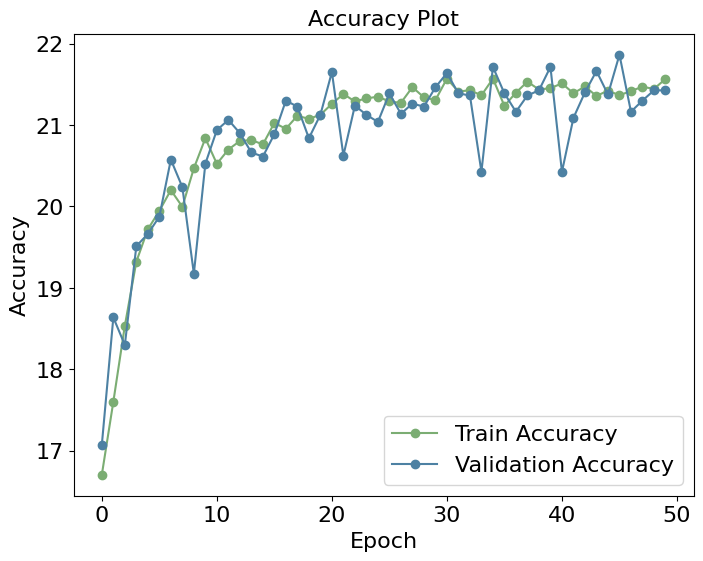}
        \caption{Training and validation loss over 50 epochs.}
        \label{fig:training_loss}
    \end{subfigure}
    \hfill
    \begin{subfigure}{0.49\linewidth}
        \centering
        \includegraphics[width=\linewidth]{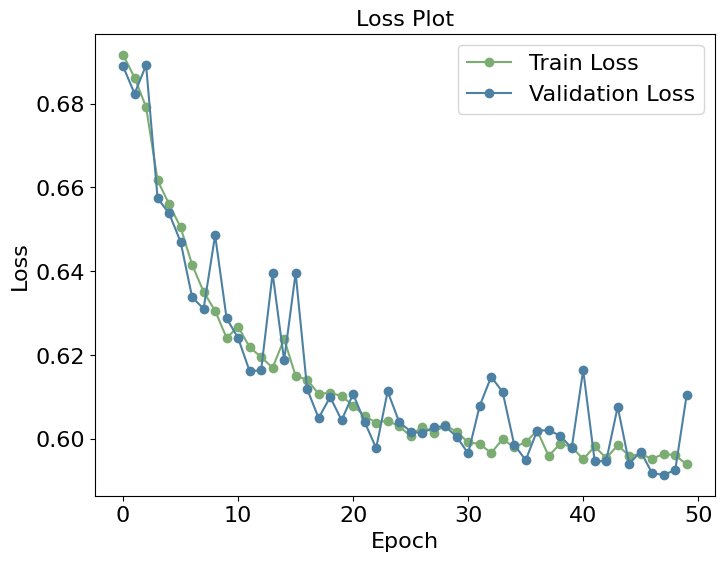}
        \caption{Training and validation accuracy over 50 epochs.}
        \label{fig:training_accuracy}
    \end{subfigure}
    \caption{Training and validation performance of Classifier 4.}
    \label{fig:training_validation_results}
\end{figure}

Classifier 4 was evaluated using a learning rate of 0.01 and trained for 50 epochs. The results of the training and validation performance are shown in Figure~\ref{fig:training_validation_results}.

The model underwent extensive hyperparameter tuning, with particular attention to the number of hidden layers, learning rate, and optimization techniques. A learning rate of 0.01 provided a stable balance between convergence and generalization. Despite this, the training and validation loss exhibited only gradual improvements over the 50 epochs. The training loss decreased from 0.6917 to 0.5833, while the validation loss declined from 0.6871 to 0.5975. Similarly, training and validation accuracy showed only modest improvements, suggesting that while the model was learning, further refinements were necessary for enhanced performance.

\begin{figure}[H]
    \centering
    \includegraphics[width=0.75\linewidth]{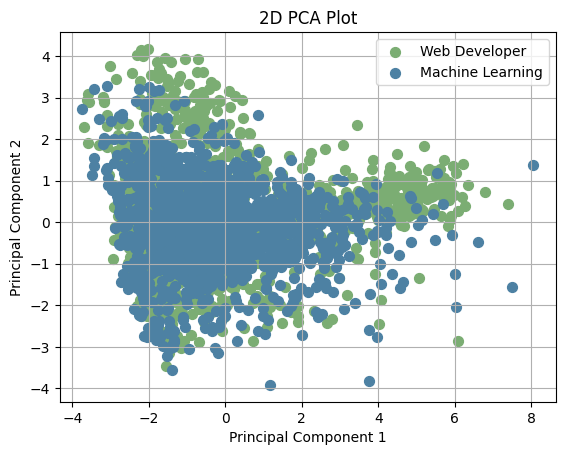}
    \caption{PCA visualization of node embeddings extracted from Classifier 4.}
    \label{fig:pca_embeddings}
\end{figure}

To better understand the learned representations, node embeddings from the final layer of Classifier 4 were extracted and reduced to two dimensions using Principal Component Analysis (PCA). The visualization in Figure~\ref{fig:pca_embeddings} indicates a significant degree of overlap between the two categories, suggesting that the learned representations may require additional refinement to improve class separability.

To further enhance classification performance, a Random Forest classifier was applied to the extracted node embeddings. This approach leveraged the high-level features learned by the GCN while introducing an ensemble learning method to improve generalization. The Random Forest classifier achieved an overall test accuracy of approximately 67\%, with a recall of 75\% and a precision of 68\% for machine learning networks. The confusion matrix revealed 841 misclassifications, with 486 errors in web development and 355 in machine learning.

\begin{figure}[H]
    \centering
    \includegraphics[width=0.75\linewidth]{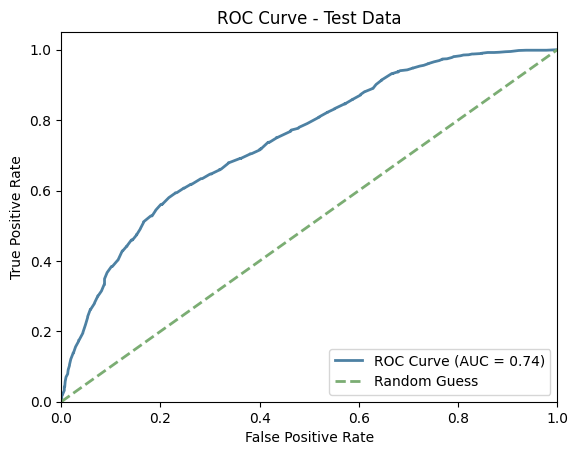}
    \caption{ROC for the Random Forest classifier applied to GCN embeddings.}
    \label{fig:roc_curve}
\end{figure}

Additionally, the AUC score was 0.74, indicating moderate classification performance. The ROC curve shown in Figure~\ref{fig:roc_curve} provides further insight into the model’s ability to distinguish between the two classes.

While these results demonstrate the feasibility of using GCNs for classifying developer networks, further improvements are necessary. Future work could explore alternative architectures, incorporate additional features, and optimize the hyperparameters of both the GCN and the Random Forest classifier to enhance performance.


\subsection{Edge-level Recommendation System}

The performance of the edge-level recommendation model was evaluated across individual network graphs using AUC scores as a primary metric (see Figure~\ref{fig:classifier_comparison}). The model was designed to predict up to five potential new connections for 10\% of randomly selected nodes within each graph. The AUC results demonstrated that the model effectively identified meaningful new connections, reinforcing its capability to capture underlying structural relationships within the networks. To further refine the recommendation system, future improvements could include implementing k-fold cross-validation to enhance robustness, incorporating more sophisticated negative sampling techniques, and evaluating additional performance metrics to provide a more comprehensive assessment of model effectiveness.

\begin{figure}[H]
    \centering
    \begin{subfigure}{0.49\linewidth}
        \centering
        \includegraphics[width=\linewidth]{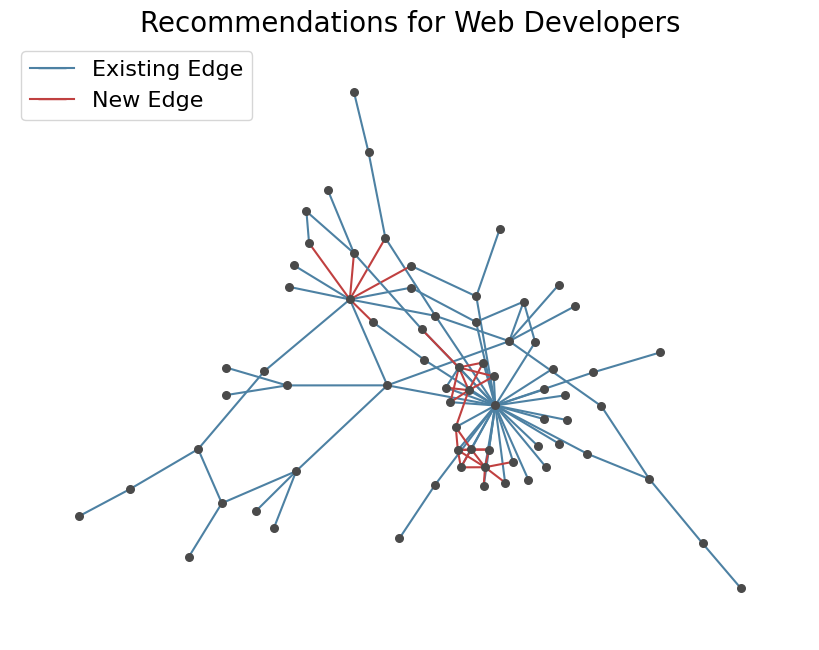}
        \caption{Web Development - Predicted Connections}
        \label{fig:webedge1}
    \end{subfigure}
    \hfill
    \begin{subfigure}{0.49\linewidth}
        \centering
        \includegraphics[width=\linewidth]{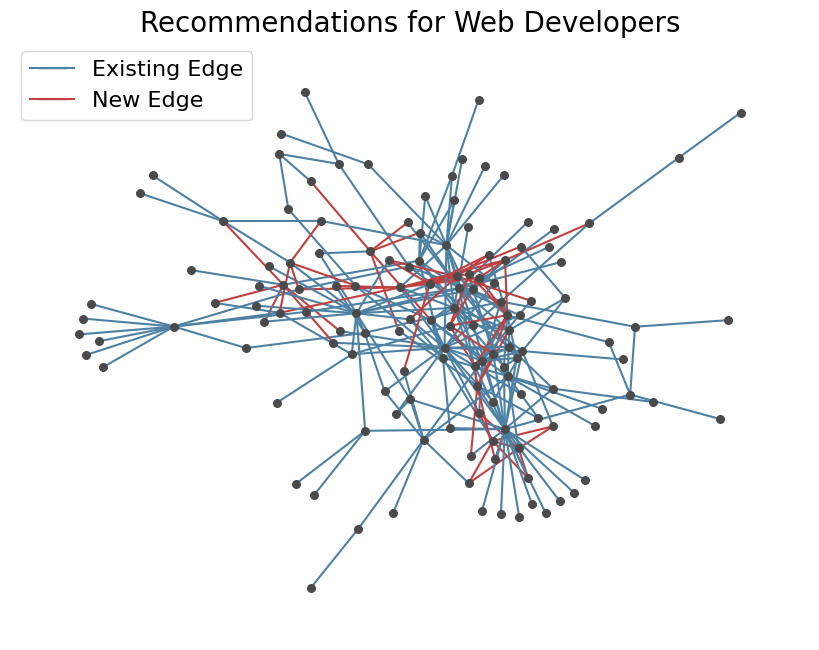}
        \caption{Web Development - Predicted Connections}
        \label{fig:webedge2}
    \end{subfigure}
    \vfill
    \begin{subfigure}{0.49\linewidth}
        \centering
        \includegraphics[width=\linewidth]{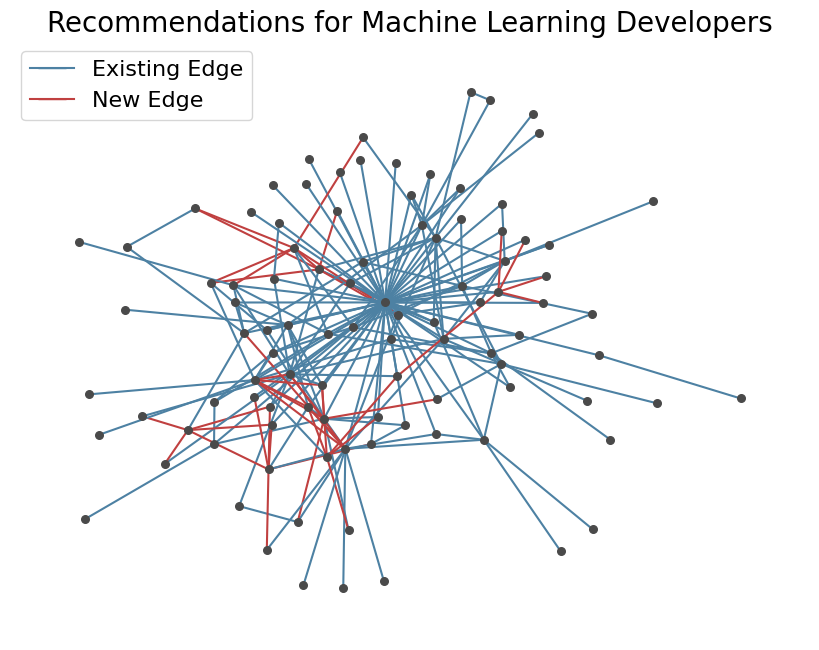}
        \caption{Machine Learning - Predicted Connections}
        \label{fig:mledge1}
    \end{subfigure}
    \hfill
    \begin{subfigure}{0.49\linewidth}
        \centering
        \includegraphics[width=\linewidth]{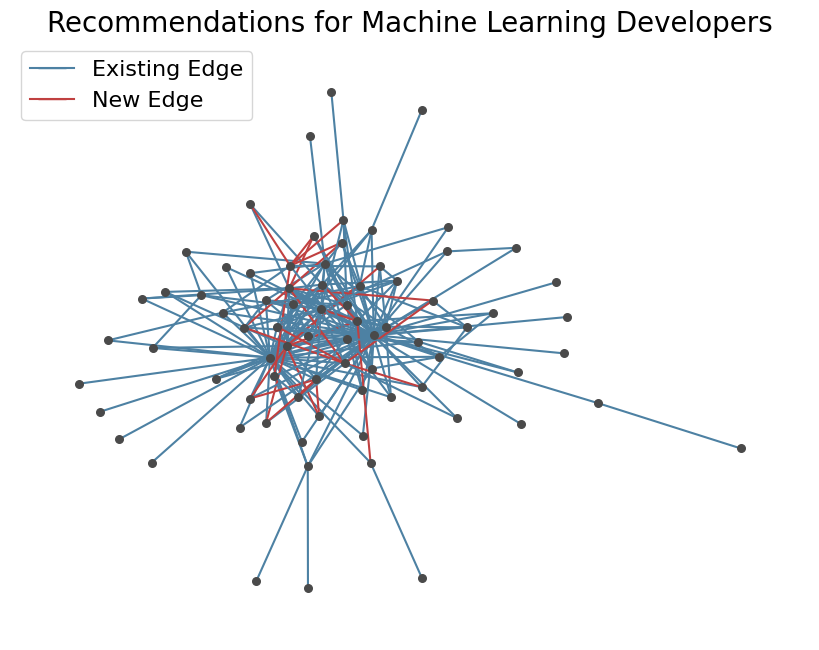}
        \caption{Machine Learning - Predicted Connection}
        \label{fig:mledge2}
    \end{subfigure}
    \caption{Visualization of recommended new connections in web development and machine learning networks.}
    \label{fig:classifier_comparison}
\end{figure}

\newpage

\section{Discussion}

The findings of this study highlight the effectiveness of graph-based methods for analyzing developer networks within the GitHub Stargazers dataset. The classification task using Graph Convolutional Networks demonstrated that network structure alone provides meaningful signals for distinguishing between web development and machine learning communities. However, the moderate accuracy and AUC scores suggest that additional node-level features, such as user activity, repository metadata, or temporal trends, could further enhance classification performance. 

Similarly, the edge-level recommendation system showed promise in identifying potential new connections within developer networks. By leveraging GraphSAGE embeddings and cosine similarity, the model successfully predicted new edges with reasonable accuracy. However, the presence of misclassified recommendations suggests that further refinement, such as more sophisticated negative sampling or the incorporation of temporal evolution, could improve the system's reliability. 

Several areas for improvement were identified. First, the classification model could benefit from additional architectural enhancements, such as attention mechanisms to weigh node influence differently based on connectivity patterns. Second, hyperparameter tuning beyond the initial grid search may further optimize model performance. Lastly, the recommendation algorithm could be extended to incorporate user interactions and collaborative filtering techniques to provide more context-aware suggestions. These refinements would make the models more robust and applicable to real-world developer community analysis.

\section{Conclusion}

This study explored graph-based approaches for classifying developer communities and predicting potential new connections within the GitHub Stargazers dataset. By utilizing GCNs for classification and GraphSAGE for recommendation, the models effectively captured the structural characteristics of developer networks. 

The classification results demonstrated that machine learning networks tend to have denser connections than web development networks, aligning with the hypothesis that communities centered around research-driven topics exhibit stronger internal connectivity. While the classification model provided valuable insights, further refinements are needed to improve its generalization capabilities. 

The recommendation system successfully identified new edges by leveraging node embeddings and similarity metrics, offering a structured way to model evolving social connections within open-source development. The findings suggest that graph-based recommendation systems can be a powerful tool for understanding how developer communities form and evolve.

Future work should explore additional feature engineering techniques, experiment with more advanced graph neural network architectures, and integrate temporal data to capture the dynamic nature of developer collaborations. Ultimately, this research provides a foundation for further advancements in social network analysis and recommendation systems within the domain of open-source software development.

\newpage
\setcounter{footnote}{1}
\bibliography{mybib} 

\begin{thebibliography}{1}
\expandafter\ifx\csname url\endcsname\relax
  \def\url#1{\texttt{#1}}\fi
\expandafter\ifx\csname urlprefix\endcsname\relax\def\urlprefix{URL }\fi
\providecommand{\bibinfo}[2]{#2}
\providecommand{\eprint}[2][]{\url{#2}}

\bibitem{carpenter2018market}
\bibinfo{author}{Carpenter, A.}
\newblock \bibinfo{title}{What is market segmentation? 5 focus areas for max roi}.
\newblock \bibinfo{howpublished}{Qualtrics} (\bibinfo{year}{2018}).
\newblock \urlprefix\url{www.qualtrics.com/experience-management/brand/what-is-market-segmentation/}.

\bibitem{karateclub}
\bibinfo{author}{Rozemberczki, B.}, \bibinfo{author}{Kiss, O.} \& \bibinfo{author}{Sarkar, R.}
\newblock \bibinfo{title}{{Karate Club: An API Oriented Open-source Python Framework for Unsupervised Learning on Graphs}}.
\newblock In \emph{\bibinfo{booktitle}{Proceedings of the 29th ACM International Conference on Information and Knowledge Management (CIKM '20)}}, \bibinfo{pages}{3125–3132} (\bibinfo{organization}{ACM}, \bibinfo{year}{2020}).

\bibitem{hagberg2008networkx}
\bibinfo{author}{Hagberg, A.~A.}, \bibinfo{author}{Schult, D.~A.} \& \bibinfo{author}{Swart, P.~J.}
\newblock \bibinfo{title}{Exploring network structure, dynamics, and function using networkx}.
\newblock In \emph{\bibinfo{booktitle}{Proceedings of the 7th Python in Science Conference (SciPy2008)}}, \bibinfo{pages}{11–15} (\bibinfo{address}{Pasadena, CA USA}, \bibinfo{year}{2008}).

\bibitem{grando2019mlcentrality}
\bibinfo{author}{Grando, F.} \emph{et~al.}
\newblock \bibinfo{title}{Machine learning in network centrality measures}.
\newblock \emph{\bibinfo{journal}{ACM Computing Surveys}} \textbf{\bibinfo{volume}{51}}, \bibinfo{pages}{1–32} (\bibinfo{year}{2019}).
\newblock \urlprefix\url{https://doi.org/10.1145/3237192}.
\newblock \bibinfo{note}{Accessed 8 Aug. 2021}.

\bibitem{dgl2023}
\bibinfo{author}{{DGL Team}}.
\newblock \bibinfo{title}{Welcome to deep graph library tutorials and documentation — dgl 0.8.2post1 documentation}.
\newblock \bibinfo{howpublished}{DGL} (\bibinfo{year}{2023}).
\newblock \urlprefix\url{https://docs.dgl.ai/}.
\newblock \bibinfo{note}{Accessed 6 Dec. 2023}.

\bibitem{zhang2018graphclassification}
\bibinfo{author}{Zhang, M.} \emph{et~al.}
\newblock \bibinfo{title}{An end-to-end deep learning architecture for graph classification}.
\newblock In \emph{\bibinfo{booktitle}{Proceedings of the AAAI Conference on Artificial Intelligence}}, vol.~\bibinfo{volume}{32} (\bibinfo{year}{2018}).
\newblock \urlprefix\url{https://doi.org/10.1609/aaai.v32i1.11782}.

\bibitem{darshan2022cosine}
\bibinfo{author}{Darshan, M.}
\newblock \bibinfo{title}{What is cosine similarity and how is it used in machine learning?}
\newblock \bibinfo{howpublished}{Analytics India Magazine} (\bibinfo{year}{2022}).
\newblock \urlprefix\url{analyticsindiamag.com/cosine-similarity-in-machine-learning/}.

\end{thebibliography}

\end{document}